\documentclass[prc,twocolumn,aps,floats,superscriptaddress,floatfix,showpacs,nofootinbib]{revtex4}
\usepackage{graphicx,amssymb,amsmath,relsize}

\def\real{\mathop{\rm Re}\nolimits}   
\usepackage{subfigure}

\begin{document}
\title{Spin flip loss in magnetic confinement of ultracold neutrons for neutron lifetime experiments}

\author{A.~Steyerl}
\email{asteyerl@uri.edu}
\affiliation{University of Rhode Island, Kingston, RI 02881, USA}
\author{K.~K.~H.~Leung}
\email{kkleung@ncsu.edu}
\affiliation{North Carolina State University, Raleigh, NC 27695, USA}
\author{C.~Kaufman}
\email{chuck@uri.edu}
\affiliation{University of Rhode Island, Kingston, RI 02881, USA}
\author{G.~M\"uller}
\email{gmuller@uri.edu}
\affiliation{University of Rhode Island, Kingston, RI 02881, USA}
\author{S.~S.~Malik}
\email{smalik@uri.edu}
\affiliation{University of Rhode Island, Kingston, RI 02881, USA}


\begin{abstract}
We analyze the spin flip loss for ultracold neutrons in magnetic bottles of the type used in experiments aiming at a precise measurement of the neutron lifetime, extending the one-dimensional field model used previously by Steyerl $\textit{et al.}$ [Phys.~Rev.~C $\mathbf{86}$, 065501 (2012)] to two dimensions for cylindrical multipole fields. We also develop a general analysis applicable to three dimensions. Here we apply it to multipole fields and to the bowl-type field configuration used for the Los Alamos UCN$\tau$ experiment. In all cases considered the spin flip loss calculated exceeds the Majorana estimate by many orders of magnitude but can be suppressed sufficiently by applying a holding field of appropriate magnitude to allow high-precision neutron lifetime measurements, provided other possible sources of systematic error are under control.  
\end{abstract}
\maketitle

\section{Introduction}\label{sec:I}

The neutron lifetime $\tau_{n}$ is an important parameter in nuclear physics, particle physics, and cosmology. $\tau_{n}$ can be combined with the neutron $\beta$-decay ($n\rightarrow p+e^{-}+\bar{\nu}_{e}$) correlation coefficients to determine the universal weak interaction vector and axial-vector coupling constants whose values allow searches for semi-leptonic scalar and tensor currents beyond the Standard Model \cite{BHA01,CIR01,GAR01}. A $\tau_{n}$ of reliable precision is also needed for calculations of the neutrino flux expected from solar and reactor sources, including detection efficiencies \cite{MEN01,ZHA01}, as well as in Big Bang nucleosynthesis calculations. At present, we are confronted by an apparent discrepancy of about three standard deviations between the average $\tau_{n}$ from ultracold neutron (UCN) storage experiments and the $\tau_{n}$ from cold neutron beam experiments. It is the leading source of uncertainty in predictions of the primordial abundance of $^{4}$He \cite{COC01,IOC01,MAT01}. For reviews of $\tau_{n}$ experiments see \cite{ABE01,DUB01,NIC01,PAU01,WIE01,SEE01}.  

One of the most promising methods to achieve higher precision for the storage-type experiments employs UCN storage in magnetic bottles. In suitable non-uniform magnetic field configurations the neutrons in one spin state should, in principle, experience no losses other than $\beta$-decay, provided that depolarization, i.e.~the non-adiabatic transition to the non-storable spin state, is sufficiently suppressed. In most existing and proposed magnetic trap systems the low-field seeking state with spin parallel to the magnetic field is being stored and a spin-flip transition to the anti-parallel state results in a loss by escape from the bottle or by interaction with the bottle walls.  

Until recently, UCN depolarization estimates \cite{VLA01, POK01} have been based on Majorana's quasi-classical result \cite{MAJ01} for a particle with spin moving with constant velocity vector through an infinitely extended non-uniform magnetic field of specific form. For field parameters as currently used for magnetic UCN storage the probability $D$ of a spin flip away from the field direction would be of order $D\sim e^{-10^{6}}$, thus immeasurably small.

Walstrom $\textit{et al.}$ \cite{WAL01}, in 2009, pointed out that the values of $D$ for confined, rather than freely moving, neutrons are much larger. For a particular vertical path in the field of the Los Alamos gravito-magnetic UCN trap they calculated $D\sim 10^{-20}-10^{-23}$, which is much larger than the Majorana estimate but still negligible in any actual or projected neutron lifetime experiment.

In \cite{STE01,STE02} we extended this theory to general orbits with both vertical and horizontal velocity components, using the model of an ideal Halbach magnetic field $\mathbf{B}$ where the magnitude $B$ only depends on the vertical position. We found that the lateral motion in the plane where the Halbach field rotates is of critical importance. Taking it into account increases the spin-flip loss thus calculated by some 10 orders of magnitude to $D\sim 10^{-12}$ for a field minimum (holding field) of $B_{h}\approx 5$ mT. This translates into a spin-flip loss rate that is a fraction $\sim\! 10^{-4}$ of the $\beta$-decay rate and decreasing rapidly with larger holding field.

The analysis in \cite{WAL01,STE01} is based on the following concepts: For the one-dimensional (1D) field model of \cite{STE01}, the potential $V(z)=gz-\mu B(z)/m=gz+|\mu|B(z)/m$ for the high-field repelled $|+\rangle$ spin state of a neutron with mass $m$ and negative magnetic moment $\mu=-60.3$ neV/T depends only on the vertical $z$ coordinate. In this model the neutrons are exposed to a uniform gravitational field $-g\hat{\mathbf{z}}$ and a non-uniform magnetic field of magnitude $B(z)$. They perform an oscillatory motion with turning points (TP) at the lower and upper horizontal equipotential surfaces (ES) where $v_{z}=0$ and the potential is $V=(E/m)-v^{2}_{\perp}/2$. Here $E$ is the neutron energy; $v_{z}$ and $v_{\perp}=\sqrt{v^{2}_x+v^{2}_{y}}$ are the vertical and horizontal velocity components, respectively. $v_{\perp}$ is constant for the 1D field model.

As the particle moves from one TP to the next, starting out in a pure $|+\rangle$ spin state, its wave component for the $|-\rangle$ spin state increases. It may change over many orders of magnitude \cite{WAL01,STE01}, peaking at critical points where the field magnitude $B$ is small and the vector $\mathbf{B}$ rotates rapidly in the reference frame of the moving particle. The spin flip probability is ``measured'' only at the next TP where, in the Copenhagen interpretation, the wave function collapses and UCNs in the $|+\rangle$ state return to the trapping region while the $|-\rangle$ projection separates in space and quickly becomes lost. Conceptually, the ``measurement'' could be made by an ideal neutron detector placed just next to the TP, which would intersect the UCNs in the ``wrong'' spin state as they exit the storage space. This ``measurement' resets the UCN wave function to a pure initial $|+\rangle$ state for the next lap where the sequence of wave evolution and collapse at the following TP is repeated.

\section{Outline}\label{sec:II}

In the present article we extend this approach to the analysis of depolarization in cylindrical multipole fields such as those described in Refs.~\cite{HUF01,EZH01,LEU01,LEU02,BEC01,MAT02}, where the trapping fields are generated by Halbach arrays of permanent magnets \cite{EZH01,LEU01,LEU02,BEC01} or, for \cite{HUF01,MAT02} and, earlier \cite{PAU02}, by superconducting currents. For these cylindrical configurations we use a 2D field model which enables us to obtain a semi-analytic expression for the ensemble-averaged spin-flip loss and which can be analyzed with no need to involve simulations. The results are consistent with the only experimental spin-flip probabilities with varying holding field available so far \cite{LEU02}.

For the cylindrical $2 N$-pole we approximate the field in cylindrical coordinates $r$, $\phi$, $\zeta$ as follows:
\begin{align}\label{1}
B_{r}&=B_{max} (r/R)^{N-1}\sin{\left(N\phi\right)},\nonumber\\
B_{\phi}&=B_{max} (r/R)^{N-1}\cos{\left(N\phi\right)},\nonumber\\
|\mathbf{B}|&=B(r)=\sqrt{B^{2}_{\zeta}+B^{2}_{max}(r/R)^{2 N-2}}\,,
\end{align}
where $N\ge 2$. $\zeta$ points along the cylinder axis and the holding field $B_{\zeta}$ is considered constant. $B_{max}$ is the trapping field magnitude at the wall and the radius $R$ (typically $\sim\! 5$ cm) is much smaller than the length, which is of order $1$ m. This justifies the neglect of gravity for horizontal configurations of this type \cite{HUF01} since the gravitational energy varies little over the trap radius. To assess the merits of model (\ref{1}) in general we have performed 3D simulations including gravity both for the vertical and the horizontal cylindrical multipole configurations.

As a second application we extend the previous analysis \cite{STE01} of depolarization for a 1D field model of the Los Alamos UCN$\tau$ trap \cite{WAL01,SAL01,SAU01} to the actual field in this magneto-gravitational trap with its asymmetrically double-curved wall in the shape of a bowl. As in \cite{WAL01} we approximate the field for the curved arrays of permanent magnets by that of the corresponding infinite planar array tangent to the bowl surface at the closest point on the bowl surface. We also use the same expressions for the flat-wall field, dubbed ``smooth'' [Eq.~(5) of \cite{WAL01}], ``one-way ripple'' [Eq.~(7)] and ``two-way ripple'' [Eq.~(8)]. The ``one-way ripple'' takes into account the finite magnet size and the ``two-way ripple'' also includes the effect of iron shims between the magnets, a design feature not implemented for the current UCN$\tau$ system (status of 2016).

The theoretical approach is outlined in Sec.~\ref{sec:III}, where we derive a first-order approximation to the spin-flip probability from the spin-dependent Schr\"odinger equation (SE), and in Sec.~\ref{sec:IV} where we average these results over the ensemble of orbits in the field configurations of UCN$\tau$ and of multipole bottles. In Sec.~\ref{sec:V} we derive a higher-order solution of the spin-dependent SE and show that it deviates very little from the first-order approximation. These various approaches are semi-classical since the field $\mathbf{B}(t)$ acting on the neutron spin is determined by the classical motion of the particle through the field. However, the results have been shown \cite{WAL01,STE01} to be consistent also with a fully quantum mechanical analysis starting from the spin and space dependent SE. We show in the Appendix that this equivalence also holds for our extension to arbitrary field configurations.   
  
As in \cite{WAL01,STE01}, we use the Wentzel-Kramers-Brillouin (WKB) approximation \cite{MOR01} to solve the SE. This is justified since the spatial variation of field variables is much slower than the variation of the UCN wave function. The scales are of order cm for gravity and $\mathbf{B}$, and of order $\mu$m or less for the wavelength. Thus the wave function for spin state $|+\rangle$ can be expressed in the WKB form except at a TP $z^{\prime}=0$, where its amplitude $1/\sqrt{k^{\prime}_{+}(z^{\prime})}$ diverges since the wave number $k^{\prime}_{+}$ vanishes. (See Eq.~(\ref{A12}) for details.)

\section{Semi-classical approach}\label{sec:III}

Neutron lifetime experiments based on magnetic storage require that the spin follows the changes of field direction along the neutron path for a time much longer than the neutron lifetime, implying that the probability $|\alpha(t)|^{2}$ for spin $|+\rangle$, parallel to the local field, is always much larger than the small spin-flipped part $|\beta(t)|^{2}$. Therefore, in order to separate large terms in the SE from the small ones it is advantageous to use a reference system which rotates with the field experienced by the moving particle. Thus we use the SE for spin 1/2 with quantization axis in the local field direction, as in \cite{WAL01,STE01}:
\begin{align}\label{2}
i\hbar \frac{d}{dt}&\left(\alpha(t) \chi^{+}(t)+\beta(t) \chi^{-}(t)\right)\nonumber\\
&=|\mu|B\left(\alpha(t) \chi^{+}(t)-\beta(t) \chi^{-}(t)\right),
\end{align}
where
\begin{align}\label{3}
&\chi^{+}=\left( \begin{array}{c}
c  \\
e_{+}s \\
 \end{array} \right)\quad\textrm{and}\quad\chi^{-}=\left( \begin{array}{c}
e_{-}s  \\
-c  \\
 \end{array} \right)
\end{align}
are the spinors aligned in the direction of or opposite to the local magnetic field $\mathbf{B}$, respectively. For the Los Alamos ``bowl'' we choose the $x$ direction (defined as the direction of the toroidal holding field $\mathbf{B}_{h}(z)$ measured in the vertical symmetry plane) as the fixed quantization axis relative to which the local field direction is given by the polar angle $\theta=\arccos(B_{x}/B)$ and the azimuthal angle $\phi=\arctan(B_{z}/B_{y})$. In (\ref{3}) we have used $c=\cos{(\theta/2)}$, $s=\sin{(\theta/2)}$ and $e_{\pm}=e^{\pm i\phi}$.
 
Differentiating $\chi^{+}(t)$ and $\chi^{-}(t)$ we get \cite{STE01}
\begin{equation}\label{4}
\dot{\chi}^{+}=A_{pp}\chi^{+}+A_{pm}\chi^{-},\,\,\,\dot{\chi}^{-}=A_{mp}\chi^{+}+A_{mm}\chi^{-},
\end{equation}
with time dependent coefficients
\begin{align}\label{5}
A_{pp}&=\frac{i}{2}\dot{\phi}(1-\cos\theta),\,\,\,A_{pm}=-\frac{1}{2}e_{+}(\dot{\theta}+i\dot{\phi}\sin\theta)\nonumber\\
A_{mp}&=-A^{\ast}_{pm},\,\,\,A_{mm}=A^{\ast}_{pp}=-A_{pp}.
\end{align}
$A_{mp}$ and $A_{pp}$ are small quantities to be treated as small perturbations.

Using Eqs.~(\ref{3})-(\ref{5}) in (\ref{2}) we obtain for the terms with $\chi^{+}$ in first order
\begin{equation}\label{6}
\dot{\alpha}+\frac{i\omega_{L}}{2}\alpha=0
\end{equation}
and for those with $\chi^{-}$
\begin{equation}\label{7}
\dot{\beta}-\frac{i\omega_{L}}{2}\beta=-\alpha A_{pm}=\frac{\alpha}{2}e_{+}(\dot{\theta}+i\dot{\phi}\sin\theta),
\end{equation}
where $\omega_{L}=2|\mu|B/\hbar$ is the local Larmor frequency. Equations (\ref{6}) and (\ref{7}) correspond to Eqs.~(57) and (58) of \cite{STE01}. The solutions of (\ref{6}) and (\ref{7}) are \cite{WAL01,STE01}, in WKB approximation,
\begin{align}
\alpha(t)&=e^{-i\Theta/2}\label{8},\\
\beta(t)&=-\frac{i A_{pm}}{\omega_{L}}e^{-i\Theta/2}=\frac{i}{2\omega_{L}}e_{+}(\dot{\theta}+i\dot{\phi}\sin\theta)e^{-i\Theta/2},\label{9}
\end{align} 
where $\Theta=\int_{0}^{t}\omega_{L}(t^{\prime})\,dt^{\prime}$ is twice the phase angle accumulated since the previous TP. Since $\alpha$ and $\beta$ have the same phase, $-\Theta/2$, the wave components $\alpha$ and $\beta$ propagate between TPs as a unit (they do not run apart). This feature had also been noted for the 1D field model \cite{STE01}.

From (\ref{9}), the probability of finding the neutron in the spin-flipped state along the way to the next TP is
\begin{equation}\label{10}
p(t)=|\beta(t)|^{2} =\frac{\dot{\theta}^{2}(t)+\dot{\phi}^{2}(t)\sin^{2}\theta(t)}{4\omega^{2}_{L}(t)}=\frac{\Omega^{2}(t)}{4\omega^2_{L}(t)}.
\end{equation}
$\Omega(t)$ is the frequency of field rotation about an axis normal to the plane defined by $\mathbf{B}$ and $\dot{\mathbf{B}}$ and can be expressed as $\Omega=|\mathbf{B}\mathbf{\times}\dot{\mathbf{B}}|/B^{2}$. This form holds since $\dot{\theta}^{2}+\dot{\phi}^{2}\sin^{2}\theta$ is the squared projection of vector $\dot{\mathbf{B}}$, as drawn from the tip of vector $\mathbf{B}$, onto the unit sphere. For any trajectory we determine the arrival time at, and the position of TPs from the condition that the velocity component along the gradient of potential $V$ vanishes, $\mathbf{v\cdot \nabla}V=0$, since, at a TP, the orbit is tangential to the ES.  

Equations (\ref{9}) and (\ref{10}) make use of an approximation which we will discuss in Sec.~\ref{sec:V} in connection with the higher-order solution given in (\ref{25}) and (\ref{27}). In short, this analysis shows the following features of $p(t)$: Starting from zero at a TP, $p(t)$ increases to the value given in (\ref{10}) within a short time of order $\mu$s. Eq.~(\ref{10}) holds over the entire remainder, typically $0.01-0.1$ s, of the motion to the next TP where $p(t)$ is ``measured''. Equation (\ref{10}) shows that the result depends only on the local field variables $\Omega(t)$ and $\omega_{L}(t)$; it is independent of the path history.

For the sequence of TPs encountered along a neutron path of total duration $T_{tot}$ we determine the spin-flip rate between consecutive TPs at $t_{i-1}$ and $t_{i}$ by dividing $p(t_{i})$, from (\ref{10}), by the time interval $\Delta T_{i}=t_{i}-t_{i-1}$. Taking into account the probability $\Delta T_{i}/T_{tot}$ of finding the particle on this path element the spin-flip rate for the entire path becomes
\begin{equation}\label{11}
1/\tau_{dep}=\frac{1}{T_{tot}}\sum_{i=1}^{n}\,p(t_{i}),
\end{equation}
where $n$ is the number of TPs encountered. Finally, the depolarization rate measured in the experiments is the ensemble average over all paths, which is determined by the source characteristics and by spectral cleaning. We assume an isotropic Maxwell spectrum, thus an energy independent phase space density (PSD). The Boltzmann factor $e^{-E/k_{B}T}$ is close to unity since UCN energies $E$, which are of order $\lesssim 10^{-7}$ eV, are much lower than $k_{B}T$ even for a low trap temperature $T$. The corresponding velocity dependence of the spectrum is $f(v)\sim v^{2}$.

Relation (\ref{10}) is also obtained in a fully quantum mechanical approach using the space and spin dependent SE in WKB approximation \cite{WAL01,STE01}. This equivalence holds for any field geometry as shown in Appendix $A$.  

\section{Ensemble average of spin-flip loss}\label{sec:IV}

In actual magnetic UCN storage systems it is difficult to make sure that the spectrum is isotropic and fills phase space uniformly up to the trapping limit. In practice this would require the complete removal of UCNs with energies slightly exceeding the limit. These tend to be in quasi-stable orbits lasting for times of the order of $\tau_{n}$, thus affecting the precision of a measurement of $\tau_{n}$. Deviations from the Maxwell spectrum may also be due to the characteristics of the UCN source and of UCN transport to the trap but we will disregard these differences since they are expected to be of minor importance for the specific loss due to spin flip.

\subsection{1D field model}\label{sec:IV.A}

For the 1D field model of Ref.~\cite{STE01} the condition of constant PSD was taken into account as follows: The spin-flip probability was averaged over the vertical velocity component $v_{z0}$ in the plane $z=z_{0}$ where the gravitational downward force, $-mg$, is balanced by the magnetic upward force, $-\mu d|\mathbf{B}|/dz$. $z_{0}$ is the 1D equivalent of an elliptic fixed point $O$ and $z=z_{0}$ is the only plane where UCNs of any energy $E$ can reside, down to $E=0$ and up to the maximum value for trapping. (Here and henceforth we set the potential $V=0$ at $O$ and assume that there are no other potential minima in the trapping region; this is the case for the field configurations presently used or proposed.) For any other height, the lower energy limit is non-zero. Therefore, to include all possible orbits in the averaging process we have to choose $z_{0}$ as the reference height, and since the statistical distribution of $v_{z0}$ values is uniform for uniform PSD, the mean depolarization rate is given as the average of $1/\tau_{dep}$ as a function of $v_{z0}$.

To analyze the depolarization between consecutive TPs we start trajectories from a TP, not from the fixed point $O$; so we have to connect the statistics at $z_{0}$, which is given by a uniform distribution of $v_{z0}$, with the distribution $P(z)$ of launching height $z$ for given $v_{z0}$. $P(z)$ follows from energy conservation: The potential at the launching point is $V(z)=v^{2}_{z0}/2$, thus $(dV/dz)dz=v_{z0}dv_{z0}$. Therefore, to represent constant spacing in velocity space ($\Delta v_{z0}=$ const.) for orbits launched at height $z$ we have to choose the density $P(z)=1/\Delta z$ as
\begin{equation}\label{12}
P(z)\propto \frac{\Delta v_{z0}}{\Delta z}=\frac{|dV(z)/dz|}{v_{z0}}\sim \frac{|dV(z)/dz|}{\sqrt{V(z)}}.
\end{equation} 

\subsection{General 3D field models}\label{sec:IV.B}

Now we adapt this result to arbitrary 3D field models as are relevant for the Los Alamos bowl \cite{WAL01}. The asymmetry introduced by the choice of two different radii of curvature along the rows of Halbach field magnets helps to randomize the orbits although fully mixing phase flow cannot be achieved. To select a representative sample of orbits for depolarization calculations we assume that most particles with energy $E$ below the trapping limit will, at some time, be found at rest. (In practice it suffices to require $v\ll v_{max}\approx 3$ m/s.) In this case the particle has just reached the ES $V=E/m=$ constant. It will then be accelerated back into the region of lower potential in the direction perpendicular to the ES.

Taking these points as the initial position for simulated trajectories starting from rest we make sure that the particle remains within the volume bounded by the ES with potential $V=E/m$ (as long as no spin flip to the $|-\rangle$ state takes place). The statistical distribution of launch points in space is determined by the following extension of (\ref{12}) to 3D geometry: We relate the launching point to the fixed point $O$ in the same way as for the 1D field model. Equating the initial energy $E=m V(x,y,z)$ with the kinetic energy at $O$, $E=m v^{2}_{0}/2$, we differentiate $V(x,y,z)=v^{2}_{0}/2$. This gives $\nabla V\,ds=v_{0}\,dv_{0}$ where $s$ is the coordinate perpendicular to the ES $V(x,y,z)=$ const.~at the launching point. Furthermore, the potential in the immediate vicinity of $O$ is constant, thus the spatial density is stationary. Hence, to satisfy uniformity of PSD the density in velocity space must be uniform: $\Delta v_{0}=$ const.,~and we have to choose the spatial density of launching points according to
\begin{equation}\label{13}
\frac{1}{\Delta s}\propto\frac{\Delta v_{0}}{\Delta s}\sim \frac{|\nabla V(x,y,z)|}{\sqrt{V(x,y,z)}}.
\end{equation}
We take the last form of (\ref{13}) as the (not normalized) probability distribution $P(x,y,z$) representing the number of launch points per unit volume at position ($x,y,z$). In the simulations we use von Neumann's acceptance or rejection method to implement this probability distribution. For this purpose we need the fraction $P/P_{max}$ where $P_{max}$ is the maximum value of $P$ for the ensemble of trajectories.

We use this method to select random launching points for simulated orbits in UCN$\tau$ and use the three field models for the Halbach array, ``smooth'', ``one-way ripple'' and ``two-way ripple'' and the toroidal-shaped holding field coil geometry described in \cite{WAL01}. Averaging the spin-flip rate (\ref{11}) over a sample of some $10^{3}$ orbits, each of duration $T_{tot}=10$ s with $n\approx 100$ TPs, for four values of holding field $B_{x0}$ at the bowl bottom, we obtain the depolarization rate $\langle 1/\tau_{dep} \rangle$ shown in Table \ref{table:one}. The results are consistent, within a factor of two, with those given in Fig.~3 of \cite{STE01} for the 1D field model: $\langle 1/\tau_{dep}\rangle\approx 4\times 10^{-9}$ s$^{-1}$ for $B_{x0}=5$ mT and $\langle 1/\tau_{dep}\rangle\approx 5\times 10^{-11}$ s$^{-1}$ for $B_{x0}=50$ mT. Fig.~\ref{fig:one} shows as squares the 1D calculation of \cite{STE01} and as circles the present 3D calculations for the smooth-field model. In the range $B_{x0}\gtrsim 5$ mT the data are represented reasonably well by the proportionality $\langle 1/\tau_{dep}\rangle\sim B^{-2}_{x0}$ as indicated by the dashed line.  

\begin{table}[tb]
\caption{{Mean depolarization rate $\langle 1/\tau_{dep}\rangle$ $[10^{-9}/s]$ for the Los Alamos UCN$\tau$ field}}
\centering
\begin{tabular}{c || c | c | c}
\hline\hline
$B_{x0} [T]$ & smooth & one-way ripple & two-way ripple \\[1ex]
\hline
$0.1$ &  $0.022(1)$ & $0.023(1)$ & $0.024(1)$ \\
$0.03$ &  $0.32(1)$ & $0.33(1)$ & $0.31(1)$ \\
$0.01$ &  $1.6(1)$ & $1.7(1)$ & $1.9(1)$ \\
$0.001$ &  $9.4(1)$ & - & - \\ [0.5ex]
\hline\hline
\end{tabular}
\label{table:one}
\end{table}

Table \ref{table:one} shows that the field ripple has a minor effect on the net depolarization loss. This is plausible since the ripple only affects the immediate vicinity, of order mm, of the wall which contributes little to the spin flip since the adiabatic condition is well satisfied in the region of high field strength near the wall.

\subsection{Cylindrical multipole fields}\label{sec:IV.C}

\subsubsection{Results for the field model of equation (\ref{1})}\label{sec:IV.C.1}

We use the field of an ideal cylindrical multipole given in Eq.~(\ref{1}), which does not take into account the deviations induced in the actual designs by the discrete geometry of electric currents \cite{HUF01,MAT02}, or by the permanent magnet blocks with constant magnetization within each block in the schemes of Refs.~\cite{EZH01,LEU01,LEU02,WAL01,BEC01}.

Equation (\ref{1}) neglects gravity and assumes a uniform holding field $B_{\zeta}$ in the axial direction. With these simplifications, the field is determined only by the polar coordinates $r$ and $\phi$ in the plane perpendicular to the axis. Moreover, the field magnitude $B$ and the force $-|\mu|\,dB/dr$ acting on a $|+\rangle$ spin UCN depend only on $r$. In this central force field the equipotential lines are concentric cylindrical shells. Energy $E$ and angular momentum $L_{\zeta}$ about the symmetry axis are conserved. (For vertical systems, $L_{z}$ is conserved also in the presence of gravity as well as for variable $B_{\zeta}$, as for end fields, as long as the potential $V$ remains cylindrically symmetric.)

The orbits of the 2D hexapole ($2 N=6$) are ellipses. Analytic expressions for the orbits, in terms of elliptic integrals, exist also for the quadrupole ($2 N=4$), decapole ($2 N=10$) and for $2 N=14$ \cite{GOL01}. Alternatively, the radial equation of motion, $\dot{r}=\sqrt{2\left(E-V_{eff}(r)\right)/m}$, in the effective potential $V_{eff}(r)=V(r)+L^{2}_{\zeta}/(2 m^{2}r^{2})$ is readily solved numerically for any $N(\ge 2)$. There are two apsidal radii, $r_{min}$ and $r_{max}$, and the orbits are symmetric about the angular positions of these TPs. Therefore, it suffices to analyze only the path section between consecutive TPs.

\begin{figure}[t]
 \vspace{0mm}\hspace*{0mm}  \includegraphics[width=75mm]{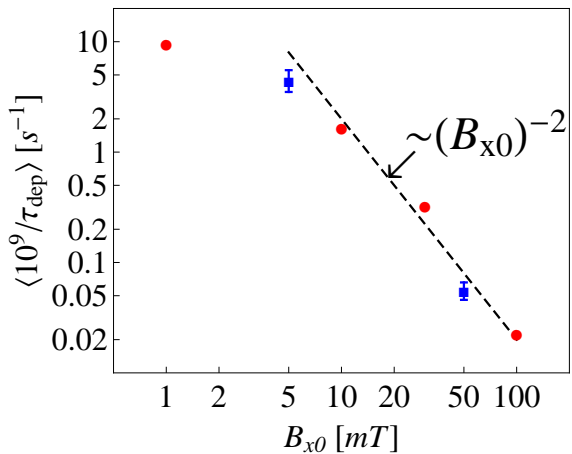}
\caption{(Color online) Comparison of spin flip loss rate calculated in \cite{STE01} for the Los Alamos UCN$\tau$ trap \cite{WAL01} using a 1D field model (blue squares) with the present 3D calculation for the smooth field given in Eq.~(5) of \cite{WAL01} (red circles). In the range $B_{x0}\gtrsim 5$ mT the data are represented reasonably well by the power law $\langle 1/\tau_{dep} \rangle\sim B^{-2}_{x0}$ as shown by the dashed line.}
  \label{fig:one}
\end{figure}

We average over all possible orbits confined within the trap radius $R$ and subject to the requirement of uniform PSD as follows. Choosing the radius $r_{1}<R$ of an ES we consider all orbits which turn around at $r_{1}$. Subset $a$ of these orbits comes from the inside and has $0\le r_{min}\le r_{1}$ and $r_{max}=r_{1}$. The other subset $b$ of orbits comes from the outside and has $r_{min}=r_{1}$ and $r_{1}\le r_{max}\le R$. In case $a$ ($b$) the region exterior to the storage space for spin $|+\rangle$ is the range $r>r_{1}$ ($r<r_{1}$). In either case, a spin-flipped UCN entering this ``forbidden zone'' is attracted toward the high field at the wall and considered as lost.

The classification $a$ or $b$ is determined by the peripheral velocity $v_{1}$ at $r_{1}$: For group $a$ the range of $v_{1}$ is between $0$ (for the radial path from or toward the center $r=0$, for which the angular momentum is zero) and $v_{c}=\sqrt{r_{1} F(r_{1})/m}$ with centripetal force $F(r_{1})=\,m dV/dr_{1}$. In the latter limit the path is circular with radius $r_{1}$. For group $b$, $v_{1}$ ranges from $v_{c}$ (circular) to $v_{2}$ for the limiting path skirting the wall ($r_{max}(v_{2})=R$). In each case, the second turning radius and the time $\Delta t(r_{1},v_{1})$ it takes from one TP to the next are found numerically from the radial equation of motion.

To determine the statistical weight of a given orbit with one TP at radius $r_{1}$ we have to modify the strategy used for the UCN$\tau$ field. In that case we considered only the sample of orbits where the particle starts from rest at the ES with the highest potential, $V=E/m$, reached for given energy $E$.

For the cylindrical multipole field (\ref{1}) only regular orbits exist and releasing a particle from rest would cover only the subset (of measure zero) of trajectories with angular momentum zero, which oscillate radially through $O$ (the axis $r=0$). However, field (\ref{1}) is an idealization and in the physical situations field irregularities such as ``ripples'' and stray fields in the axial $\zeta$ direction  are unavoidable. As far as the statistics of orbits perturbed in this way goes, the following strategy appears justified: For a path turning around at $r_{1}$ with peripheral velocity $v_{1}$ we consider the ES of radius $\rho$ such that $V(\rho)=E/m=V(r_{1})+v_{1}^{2}/2$ and relate the statistical weight for radius $\rho$ to the uniform phase-space density at $O$. As in Secs.~\ref{sec:IV.A} and \ref{sec:IV.B} we differentiate the energy balance between $r=\rho$ and $r=0$, $E/m= V(\rho)=v^{2}_{0}/2$, and obtain the proportionality
\begin{equation}\label{14}
\frac{\Delta v_{0}}{\Delta \rho}\sim \frac{dV(\rho)/d\rho}{\sqrt{V(\rho)}}.
\end{equation}
Multiplying by $\rho$ to take into account the number of allowed points along the circle with radius $\rho$ we derive for the weighting factor for radius $\rho$, and therefore also for the probability $P(r_{1},v_{1})$ for an orbit with apsidal radius $r_{1}$ and apsidal velocity $v_{1}$:  
\begin{equation}\label{15}
P(r_{1},v_{1})=\rho\,\frac{dV(\rho)/d\rho}{\sqrt{V(\rho)}}
\end{equation}
where, by definition of $V(\rho)$ and using (\ref{1}),
\begin{align}\label{16}
V(\rho)&=V(r_{1})+v_{1}^{2}/2\nonumber\\
&=(|\mu|/m)\left(\sqrt{B^{2}_{\zeta}+B^{2}_{max}(\rho/R)^{2 N-2}}-B_{\zeta}\right).
\end{align}
To evaluate $dV(\rho)/d\rho$ in (\ref{15}) we have to take into account the dependence of $V(\rho)$ and of
\begin{equation}\label{17}
\rho=R\left\{\frac{[m V(r_{1})+|\mu|B_{\zeta}\,+m v_{1}^{2}/2]^{2}-|\mu|^{2}B^{2}_{\zeta}}{|\mu|^{2}B^{2}_{max}} \right\}^{1/(2 N - 2)}
\end{equation}
(from (\ref{16})) on $r_{1}$ and $v_{1}$:
\begin{equation}\label{18}
\frac{dV(\rho)}{d\rho}=\frac{\partial V(\rho)/\partial r_{1}}{\partial \rho/\partial r_{1}}+\frac{\partial V(\rho)/\partial v_{1}}{\partial \rho/\partial v_{1}}=2 \frac{dV(r_{1})/dr_{1}}{\partial \rho/\partial r_{1}},
\end{equation}
where we have used $\partial \rho/\partial v_{1}=[v_{1}/(dV(r_{1})/dr_{1})]$ $\times (\partial \rho/\partial r_{1})$ which follows from (\ref{17}) with the help of $\partial V(\rho)/\partial r_{1}=dV(r_{1})/dr_{1}$ and $\partial V(\rho)/\partial v_{1}=v_{1}$. The result is
\begin{align}\label{19}
&P(r_{1},v_{1})\sim\nonumber\\
&\frac{\mu^{2}B^{2}_{t}(r_{1})+|\mu| B(r_{1}) m v_{1}^{2}+m^{2}v_{1}^{4}/4}{\left(|\mu|B(r_{1})+m v_{1}^{2}/2\right)\sqrt{|\mu | [B(r_{1})-B_{\zeta}]+m v_{1}^{2}/2}},
\end{align}
where $B_{t}(r)=B_{max}(r/R)^{2 N - 2}$ and $B(r)=\sqrt{B_{\zeta}^{2}+B^{2}_{t}}$ are the multipole fields without and with holding field $B_{\zeta}$, respectively.

Weight factor (\ref{19}) determines how the depolarization rate from (\ref{10}) is averaged over all paths. At TPs we have $\dot{\theta}=0$ since $\theta$ depends only on $r$ and $\dot{r}=0$. (Here we measure $\theta$ from the $\zeta$-axis.) The angular velocity of trapping field rotation experienced by a neutron moving through a TP is $\dot{\phi}=(N-1)v_{1}/r_{1}$. Thus, averaging $p/\Delta T$ from (\ref{10}) over the ensemble of paths the overall depolarization rate becomes
\begin{align}\label{20}
\left\langle 1/\tau_{dep}\right\rangle=&\frac{(N-1)^{2}}{\nu}\int_{r_{1}=0}^{R}dr_{1}\frac{\sin^{2}\theta(r_{1})}{4 r^{2}_{1}\omega^{2}_{L}(r_{1})}\nonumber\\
&\times\int_{v_{1}=0}^{v_{2}(r_{1})}\,dv_{1}\,P(r_{1},v_{1})\frac{v_{1}^{2}}{\Delta t(r_{1},v_{1})},
\end{align}
with Larmor frequency $\omega_{L}(r_{1})$ at radius $r_{1}$ and normalization constant $\nu=\int_{0}^{R}dr_{1}\int_{0}^{v_{2}(r_{1})}dv_{1}\,P(r_{1},v_{1})$.

\begin{figure}[t!]
 \includegraphics[width=85mm]{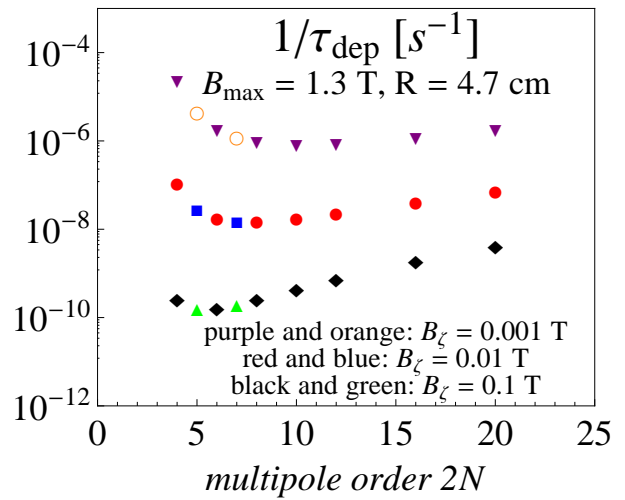}
\caption{(Color online) Mean spin-flip rates calculated from Eq.~(\ref{20}) for a cylindrical multipole trap vs.~order $2N$. The purple, red and black points (down triangles, closed circles and diamond symbols) represent integer $N$; the intervening orange, blue and green points (open circles, squares and up triangles) are for half-integral $N$.}
  \label{fig:two}
\end{figure}

Numerical results for a wide range of multipole orders $2N$ are shown in Fig.~\ref{fig:two} for $R=4.7$ cm and $B_{max}=1.3$ T. These are typical values for multipole traps; we keep these parameters the same for hypothetical systems where only the multipole order $2 N$ is varied and use $B_{\zeta}=0.001$ T, $0.01$ T and $0.1$ T for the holding field. To show the behavior of $\langle 1/\tau_{dep}\rangle$ vs.~$N$ more clearly we have added the half-integral values $N=5/2$ and $N=7/2$ which cannot be realized with magnetic fields.

Our calculation for the octupole ($2N=8$) at $B_{\zeta}=1$ mT gives $\langle\tau^{-1}_{dep}\rangle=1.5\times 10^{-5}$ s$^{-1}$. This result is consistent with the order of magnitude $\tau_{dep}=(4\pm 16)\times 10^{4}$ s$^{-1}$ given in Table IV of \cite{LEU02} for a solenoid current $3$ A which corresponds to $B_{\zeta}\gtrsim 1$ mT. (In experiment \cite{LEU02} some depolarization may have been caused by reflection on the Fomblin-coated wall at the bottom of the trap.)

The depolarization rates calculated from (\ref{20}) for the multipole traps are about $10^{2}$ times those for the 3D UCN$\tau$ field for the same holding field. The difference may be attributed to the small radius $R=4.7$ cm used. The dimensions of the UCN$\tau$ field are larger, $\sim\!0.5$ m, and therefore the average field gradient is smaller. Our calculations for a cylindrical multipole with large radius $R=1$ m, a value similar to the multipole design of Ref.~\cite{MAT02}, gives $\approx 10^{2}$ times lower spin-flip losses for the same values of $B_{\zeta}$.

We have evaluated expression (\ref{20}) for the mean depolarization rate, taking into account all possible flight paths subject to the condition of constant PSD and confined to a cylinder of radius $R$. This was possible since all orbits are regular for model field (\ref{1}).

By contrast, the orbits in the actual magnetic traps are perturbed and may show instability. In this case we rely on sampling. For instance, for the Los Alamos UCN$\tau$ system with its field asymmetry we have considered, in Sec.~\ref{sec:IV.B}, only orbits for which the particle velocity and angular momentum vanish at some time.

Applying the same method to 3D simulations for vertical multipole configurations including gravity would not provide a proper sample of orbits since these systems conserve angular momentum about the vertical axis, $L_{z}=0$. All paths launched from rest would be confined to vertical planes passing through the central axis, as stated earlier. 
 
\subsubsection{3D simulations for multipole fields}\label{sec:IV.C.2}

We include orbits with $L_{z}\neq 0$ as follows. Choosing a random initial position $Q$ within the trap volume a particle is launched with initial velocity vector $\mathbf{v}_{1}$ tangential to the ES at $Q$ and pointing in a random direction within the launch plane. To conform to a uniform distribution in velocity space the endpoint of $\mathbf{v}_{1}$ is uniformly distributed within the area of a disk whose radius is determined by the maximum velocity for particle trajectories confined to the trap volume. This and the following operations correspond to those described for the 2D field model (\ref{1}) in Sec.~\ref{sec:IV.C.1} but averages over the circular ESs of the latter model are now replaced by averages over ESs of general shape in 3D space. Based on (\ref{13}), this leads to an approximation for the weight factor $P(\mathbf{r}_{1},\mathbf{v}_{1})$ for launch at position $\mathbf{r}_{1}$ and initial velocity $\mathbf{v}_{1}$ and, finally, to the mean depolarization rate $\langle 1/\tau_{dep} \rangle$ by averaging (\ref{11}) over some $10^{3}$ orbits, each of duration $T_{tot}=10$ s with $n\gtrsim 500$ TPs.

\begin{table}[tb]
\caption{Mean depolarization rate $\langle 1/\tau_{dep}\rangle$ $[10^{-9}/s]$ from 3D simulations for cylindrical multipoles with gravity and end coils included}
\centering
\begin{tabular}{c || c | c | c}
\hline\hline
$2N$ & 4 & 8 & 20 \\[1ex]
\hline
HOPE, vertical &  $[4.1(1)]$ & $0.34(1)$ & $[0.78(1)]$  \\
HOPE, horizontal &  $[0.61(1)]$ & $0.15(1)$ & $[0.89(1)]$ \\
NIST, mark 2 &  $0.43(1)$ &  &  \\
NIST, mark 3 &  $0.021(1)$ &  &  \\ [0.5ex]
\hline\hline
\end{tabular}
\label{table:two}
\end{table}

We approximate the effect of spectral cleaning in UCN storage experiments by specifying the largest energy $E_{max}$ of stored particles. The calculations of $\langle 1/\tau_{dep} \rangle$ shown in Table \ref{table:two} use $E_{max}\approx 0.8$ times the value $|\mu|B_{high}$ for the highest field $B_{high}$ in the trap and are based on the field parameters of the following two designs:

(a) The HOPE octupole magnet \cite{LEU02} has bore radius $4.7$ cm and its axis oriented vertically or horizontally. For the vertical configuration we assume activation of only the bottom solenoid with a maximum axial field of $1.4$ T while gravity provides the cap. With both end field solenoids activated in the horizontal configuration we assume fields of $1.4$ T on both ends, separated by a distance of $1.13$ m, without activating the long holding field solenoid. The radial confinement field is $B_{max}=1.3$ T at the trap wall.

(b) Two horizontal Ioffe type quadrupole magnets have been used at NIST \cite{BRO01,YAN01} (versions mark 2 and mark 3). For mark 2 (mark 3) we use maximal fields of, axially: $1.4$ T ($4.0$ T), and radially: $1.3$ T ($3.9$ T), a bore radius of $5$ cm ($5$ cm) and a separation of $0.4$ m ($0.76$ m) between the centers of the end field solenoids; for mark 2, the latter include ``bucking'' coils \cite{BRO01,YAN01} causing the axial field to drop off more quickly to a minimum of $0.1$ T at the trap center. For mark 3, the minimum field is $0.6$ T.

Since the depolarization rates depend only weakly on the details of the field distribution, such as field ripples, we use the smoothed fields and approximate the solenoid fields by their values on the solenoid axis, neglecting the variations of field magnitude and direction over the bore cross section. The results are shown in Table \ref{table:two}.

For the HOPE-type system we include, in square brackets, also systems with the same geometries and maximum fields but different multipole order $2 N$. We observe the same tendency as for the 2D calculations of Fig.~\ref{fig:two}: The depolarization rates for $2 N=4$ and $2 N=20$ are higher than in the intermediate range ($2 N\sim 8$). This can be explained by higher field gradients, near the axis for low $N$ and near the wall for high $N$.

As a further general feature, the magnitudes and $N$ dependence of $\langle 1/\tau_{dep}\rangle$ from the 3D simulations in Table \ref{table:two} are well approximated by the 2D results from Fig.~\ref{fig:two} if we use values of holding field close to their minima: $0.08$ T/$0.1$ T for HOPE (vertical configuration with $B_{min}=0.013$ T/horizontal with $B_{min}=0.09$ T) \cite{LEU02} and $0.1$ T ($0.6$ T) for NIST mark 2 (mark 3) with $B_{min}\approx 0.1$ T ($0.6$ T) \cite{BRO01,YAN01}.

Finally, $\langle 1/\tau_{dep}\rangle$ approximately scales like $B^{-2}_{\zeta}$. A similar increase of $\langle 1/\tau_{dep} \rangle$ with decreasing holding field is also seen for UCN$\tau$ as shown by the dashed line in Fig.~\ref{fig:one}. As an application of scaling in an experiment, we could deliberately lower the holding field to enhance the depolarization loss to a measurable level to verify that the loss for the actual field is negligible. 

\section{A higher-order solution}\label{sec:V}

In this section we compare the first-order approximation for the spin-dependent SE, Eqs.~(\ref{6}), (\ref{7}), with a higher-order approach where we retain all the terms with $A_{pm}$ and $A_{pp}$ [given in (\ref{5})], which arose from the transformation to the reference system rotating with the field:
\begin{align}
&\dot{\alpha}+\frac{i\omega_{L}}{2}\alpha=-A_{pp} \alpha+A^{\ast}_{pm}\beta,\label{21}\\
&\dot{\beta}-\frac{i\omega_{L}}{2}\beta=-A_{pm} \alpha+A_{pp}\beta.\label{22}
\end{align}  
The coupled first-order ODEs (\ref{21}) and (\ref{22}) can be solved by direct numerical integration with initial conditions $\alpha(0)=1$, $\beta(0)=0$ for a particle starting in the $|+\rangle$ state at $t=0$.

Alternatively, we can use the perturbation approach developed in \cite{STE03} for searches for a permanent electric dipole (EDM) of the neutron, to solve the SE for spin $1/2$ up to second order of small perturbations. In the EDM case the UCN spin state is perturbed by magnetic field inhomogeneities and a strong static electric field. In Eqs.~(\ref{21}) and (\ref{22}) the perturbations are the terms on the RHS, which are much smaller than those on the left.

To facilitate comparison with \cite{STE03} we define $\Sigma_{pp}(t)=-2 i A_{pp}(t)$ (real-valued), $\Sigma(t)=-2 i A_{pm}(t)$ (complex), $\omega_{1}(t)=\omega_{L}(t)+\Sigma_{pp}(t)$ and $\Theta_{1}(t)=\int_{0}^{t}\omega_{1}(t^{\prime})\,dt^{\prime}$. In practical cases, $\Sigma_{pp}$ is at least $10^{4}$ times smaller than $\omega_{L}$; thus $\omega_{1}$ is very close to $\omega_{L}$.

The transformations $\alpha(t)=u(t)\,e^{-i\Theta_{1}(t)/2}$, $\beta(t)=w(t)\,e^{i\Theta_{1}(t)/2}$ turn Eqs.~(\ref{21}), (\ref{22}) into
\begin{align}\label{23}
&i\dot{u}(t)=\frac{1}{2}\Sigma^{\ast}(t)\,w(t)\,e^{i\Theta_{1}(t)}\textrm{,}\nonumber\\
&i\dot{w}(t)=\frac{1}{2}\Sigma(t)\,u(t)\,e^{-i\Theta_{1}(t)}.
\end{align}
These coupled ODEs for $u$ and $w$ have the same form as Eqs.~(7) for $\alpha_r$ and $\beta_{r}$ in \cite{STE03}. The only difference is the arguments of the exponential functions. In \cite{STE03}, the SE was transformed into the reference frame rotating at constant frequency $\omega_{0}$ for constant applied Larmor field. In the present case, $\omega_{1}(t)$ can be an arbitrary function of $t$; thus the phase factor $e^{\pm i\omega_{0}t}$ is replaced by $e^{\pm i\Theta_{1}(t)}$. 

We combine the two first-order ODEs (\ref{23}) into the single second-order ODE for $u(t)$:
\begin{equation}\label{24}
\ddot{u}(t)-\left(i\omega_{1}(t)+\frac{\dot{\Sigma}^{\ast}(t)}{\Sigma^{\ast}(t)}\right)\dot{u}(t)=-\frac{1}{4}|\Sigma(t)|^{2}u(t)
\end{equation} 
which has the same form as Eq.~(8) of \cite{STE03} with $\omega_{0}$ in the first term of the expression in brackets replaced by $\omega_{1}(t)$. This additional time dependence does not affect the method of solving (\ref{24}) since the second term is time-dependent in either case. The initial conditions, $u(0)=1$ and $w(0)=0$, are the same as for the EDM case with initial spin up ($\alpha_{r}(0)=1$, $\beta_{r}(0)=0$).

Following the steps (11) to (18) of \cite{STE03} we derive
\begin{align}\label{25}
w(t)&=\frac{2 i \dot{u}(t)}{\Sigma^{\ast}(t)}\,e^{-i \Theta_{1}(t)}=-\frac{i}{2}\left(\Sigma_{i}(t)-\Sigma_{i}(0) \right)\nonumber\\
&=-\frac{i}{2}\int_{0}^{t}dt^{\prime}\,e^{-i\Theta_{1}(t^{\prime})}\,\Sigma(t^{\prime})
\end{align} 
where $\Sigma_{i}(t)=\int dt\,e^{-i\Theta_{1}(t)}\,\Sigma(t)$.

The initial value, $w(0)=0$ at $t=0$, satisfies the required initial condition $\beta(0)=0$. As $t$ increases, the integral in (\ref{25}) rapidly increases on a time scale of order $t_{min}=1/\omega_{1}(0)\approx 1/\omega_{L}(0)$. In practical application this is a very short time since the Larmor frequency $\omega_{L}$ is large everywhere inside the trap volume, even at the field minimum where $B$ is the holding field. For $B(0)=0.001$ T, $t_{min}=\pi\hbar/(|\mu|B(0))\approx 30$ $\mu$s.

For times $t\gg t_{min}$ it is advantageous to change the integration variable in (\ref{25}) from $t^{\prime}$ to $\Theta_{1}$, with $dt^{\prime}=d\Theta_{1}/\omega_{1}$, and to integrate by parts:
\begin{align}\label{26}
w(t)=&\left[\frac{\Sigma(t)\,e^{-i\Theta_{1}(t)}}{2\omega_{1}(t)} \right]^t_{0}-\frac{1}{2}\int_{0}^{t}\,dt^{\prime}\,e^{-i\Theta_{1}(t^{\prime})}\frac{d}{dt^{\prime}}\left[\frac{\Sigma(t^{\prime})}{\omega_{1}(t^{\prime})} \right].
\end{align}
We did not employ the WKB approximation to derive Eq.~(\ref{26}) but its use enables us to evaluate it analytically: We can neglect the last term in (\ref{26}) since the field variable $\Sigma/\omega_{1}$ varies slowly on the wavelength scale and get, with (\ref{5}),
\begin{align}\label{27}
\beta(t)&=w(t)\,e^{i\Theta_{1}(t)/2}=\frac{\Sigma(t)\,e^{-i\Theta_{1}(t)/2}}{2 \omega_{1}(t)}\nonumber\\
&=\frac{i}{2\omega_{1}(t)}e_{+}(\dot{\theta}+i\dot{\phi}\sin\theta)e^{-i\Theta_{1}(t)/2}.
\end{align}

In (\ref{27}) we have set the integration constant from the lower limit $t=0$ of the first term in (\ref{26}) equal to zero, and an equivalent approximation had also been made in deriving the first-order solution (\ref{9}) which differs from (\ref{27}) only by the replacement of $\omega_{1}$ by $\omega_{L}$. The detailed justification in \cite{STE01}, below Eq.~(28), can be summarized as follows. Equations (\ref{9}) and (\ref{27}) are semi-classical since the SE is solved for the time dependent field $\mathbf{B}(t)$ determined from the classical equations of motion. A fully quantum mechanical treatment requires the solution of the spin and space dependent SE as in \cite{WAL01,STE01} and in Appendix A. In this quantum analysis the exact solution for the wave function near a TP involves the Airy functions and the TP is blurred into a non-zero region (typically of order $\mu$m). So is the starting time at a TP. Thus, in (\ref{26}) the initial value of $w(t)$ at $t_{0}\approx 0$, which is $\sim\! e^{-i \Theta_{1}(t_{0})}\approx e^{-i\omega_{1}(0)t_{0}}$, should be averaged over the rapidly varying phase $\omega_{1}(0)t_{0}$ with the result $w(0)=\left\langle e^{-i\Theta_{1}(t_{0})}\right\rangle_{t_{0}} =0$. This holds except for the initial few micrometers of the path subsequent to a TP.

Outside of this region we have, from (\ref{27}) and (\ref{5}),
\begin{equation}\label{28}
|\beta(t)|^{2}=|w(t)|^{2}=\frac{|\Sigma(t)|^{2}}{4\omega^{2}_{1}(t)}=\frac{|A_{pm}(t)|^{2}}{\omega^{2}_{1}(t)}=\frac{\Omega^{2}(t)}{4 \omega^{2}_{1}(t)}.
\end{equation}  
Eq.~(\ref{28}) agrees with the first-order solution (\ref{10}) if we replace $\omega_{L}$ by $\omega_{1}$. As we have seen, in practical cases the difference between $\omega_{L}$ and $\omega_{1}$ is negligible. 

\begin{figure}[t]
 \vspace{0mm}\hspace*{0mm}  \includegraphics[width=85mm]{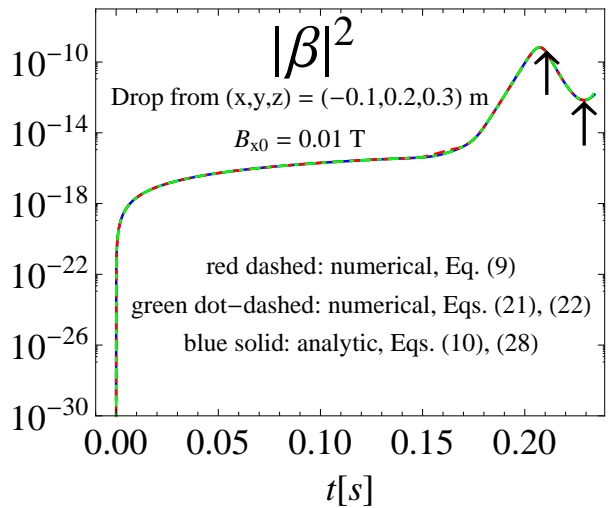}
\caption{(Color online) Magnitude squared of spin-flip amplitude $\beta$ for UCNs released from rest at position $(x,y,z)=(-0.1,0.2,0.3)$ m in the Los Alamos UCN$\tau$ ``smooth field'' \cite{WAL01}. The analytic results, Eqs.~(\ref{10}) and (\ref{28}) (blue solid), and the numerical solutions of ODE (\ref{9}) (red dashed) and of ODEs (\ref{21}), (\ref{22}) (green dot-dashed) closely agree. The first two turning points, shown by arrows, are at $0.2108$ s (when the particles pass the field minimum at a close distance) and $0.2291$ s (when they are reflected at the high field near the wall). The reset of $\beta$ to zero at these turning points and its subsequent fast increase, within microseconds or less, to the values given by the curves are not shown.}
  \label{fig:three}
\end{figure}

\begin{figure}[tbh]
 \vspace{0mm}\hspace*{0mm}  \includegraphics[width=85mm]{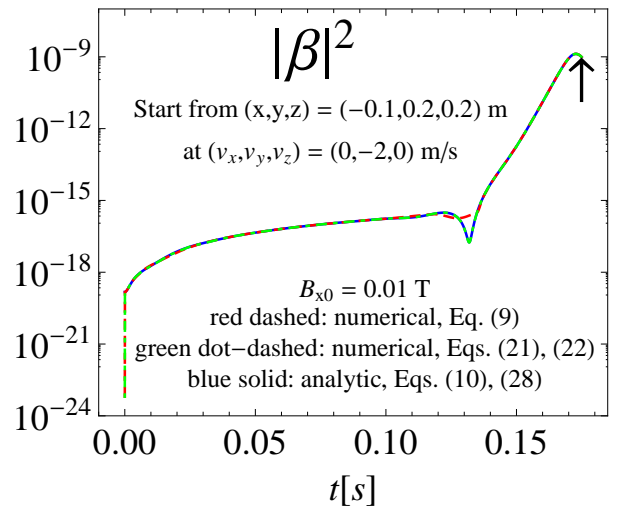}
\caption{(Color online) Spin-flip probability $|\beta|^{2}$ for UCNs launched in UCN$\tau$ at time $t=0$ from a TP at $(x,y,z)=(-0.1,0.2,0.2)$ m at velocity $(v_{x},v_{y},v_{z})=(0,-2,0)$ m/s (tangential to the local equipotential surface). The numerical integrations of ODE (\ref{9}) (red dashed) and of ODEs (\ref{21}), (\ref{22}) (green dotted) show the rapid increase, within a few Larmor periods, from $\beta=0$ at $t=0$ to the asymptotic behavior given by Eqs.~(\ref{10}) and (\ref{28}) (blue solid). The ensuing evolution of $|\beta|^{2}$ up to the next TP (shown by the arrow) is practically unaffected by the transient behavior at $t\approx 0$.}
  \label{fig:four}
\end{figure}
Fig.~\ref{fig:three} shows the time-dependence of spin-flip probability $|\beta(t)|^{2}$ for a particle released from rest at an arbitrary position in UCN$\tau$, here $(x,y,z)=(-0.1,0.2,0.3)$ m, calculated in three ways: (a) The analytic results from Eqs.~(\ref{10}) and (\ref{28}) (solid curve), and the numerical solutions (b) of differential Eq.~(\ref{9}) (dashed) and (c) of ODEs (\ref{21}) and (\ref{22}) (dot-dashed). The three curves closely agree except within short time intervals $\delta t \lesssim 10$ $\mu$s subsequent to passage though TPs at non-zero velocity, such as the TPs marked by the arrows. The deviations (not shown in Fig.~\ref{fig:three}) are due to the reset to $\beta=0$ at a TP. They are shown in detail in Fig.~\ref{fig:four} for a UCN moving away from a TP at $(x,y,z)=(-0.1,0.2,0.2)$ m at initial velocity $2$ m/s tangential to the local ES. As expected from Eq.~(\ref{26}), $|\beta|^{2}$ jumps, within a few Larmor periods ($\lesssim 10$ $\mu$s), from $0$ to the asymptotic curve given by (\ref{10}) (or (\ref{28})). The remainder of the wave evolution up to the next TP, shown by the arrow, is practically unaffected by the transient at $t\approx 0$.

As a crucial test of the validity of numerical integration of (\ref{21}), (\ref{22}) we verified the norm, $|\alpha(t)|^{2}+|\beta(t)|^{2}$, to be $1$ within $1$ ppm. The demands on the precision of numerical integration of (\ref{9}), (\ref{21}) and (\ref{22}) become more stringent for paths through regions of higher magnetic field since large Larmor frequencies require short time steps.

\section{Summary and conclusions}\label{sec:VI}

The spin-flip loss in magnetic storage of UCNs in the Los Alamos UCN$\tau$ permanent magnet trap had been analyzed theoretically in \cite{WAL01} for neutrons on a specific vertical path, and in \cite{STE01} for arbitrary motion. In the latter work we used a 1D model for the trapping field. In the present article we have extended this analysis to arbitrary orbits in arbitrary fields in 3D space and report calculations of mean spin-flip rates for the UCN$\tau$ system and for multipole fields such as the cylindrical octupole of the HOPE project \cite{LEU01,LEU02} and the Ioffe-type quadrupole trapping fields of \cite{HUF01,BRO01,YAN01}. We have also investigated a simplified 2D field model for the cylindrical multipole fields and shown that it yields analytic results for depolarization probabilities which are consistent with the more elaborate 3D simulations. In all cases relevant to magnetic UCN storage we have established agreement between the semi-classical approach, solving the spin-dependent SE for the time-dependent field seen by the particle in a classical orbit, and a fully quantum mechanical analysis based on the space and time dependent SE solved in WKB approximation. The relative difference between a first-order treatment (in Sec.~\ref{sec:III}) and a higher-order analysis (in Sec.~\ref{sec:V}) of depolarization in the semi-classical framework is at most on the order of $10^{-4}$ in practical applications.  

We confirm and generalize the earlier conclusions of \cite{WAL01,STE01} relating to ``Majorana spin flip at zeros of the magnetic field''. Magnetic UCN traps avoid locations of vanishing field by applying a holding field $B_{h}$ perpendicular to the trapping field. For typical values of $B_{h}$ we calculate spin-flip probabilities which are greater, by many orders of magnitude, than the Majorana prediction \cite{MAJ01} which had been derived for an infinitely extended field rather than trapping fields of finite extent. For the magnetic traps investigated we have found an approximate power law behavior of spin-flip loss rate as a function of $B_{h}$: $1/\tau_{dep}\sim B^{-2}_{h}$ (to be compared with the exponential behavior, $\sim e^{-\pi\xi/2}$ with adiabaticity parameter $\xi=\omega_{L}/\Omega$, for the Majorana field model \cite{MAJ01}). This implies that $B_{h}$ can be made large enough to render spin-flip loss negligible as compared to other possible sources of systematic error in neutron lifetime experiments, foremost that due to marginal trapping (leaving aside the more fundamental question raised in \cite{GRE01} whether or not neutron lifetime values derived from storage experiments should indeed be identical to those from beam-type experiments). We have shown that this conclusion is not restricted to the simplified 1D field model previously used in \cite{STE01} but holds also for the actual 3D fields in suitable magnetic neutron bottles. 

\acknowledgments

We are grateful to P.~Huffman for very helpful discussions.

\appendix

\section{Quantum analysis}\label{sec:A}

The space and spin dependent wave function $\Psi$ for a neutron with energy $E$ moving in a gravito-magnetic trapping field satisfies the SE
\begin{equation}\label{A1}
E\Psi=\left[-\frac{\hbar^{2}}{2m}\nabla^{2}+m g z+|\mu|\boldsymbol{\sigma \cdot B}(x,y,z) \right] \Psi
\end{equation}
where $\Psi=\alpha^{(3)}(x,y,z)\chi^{+}+\beta^{(3)}(x,y,z)\chi^{-}$ and $\sigma_{x}$, $\sigma_{y}$ and $\sigma_{z}$ are Pauli matrices. Superscript $(3)$ indicates that $\alpha^{(3)}(x,y,z)$, for the spin-up wave (relative to the local magnetic field direction), and $\beta^{(3)}(x,y,z)$, for the spin-down wave, are functions of the three spatial coordinates.  

The derivatives of $\chi^{+}$ and $\chi^{-}$ with respect to $j=x,y,z$ are of the same form as the temporal derivatives (\ref{4}). In terms of the spin angles $\theta$ and $\phi$ and of $e_{\pm}=e^{\pm i\phi}$ as defined below Eq.~(\ref{3}) we have
\begin{align}
&\chi^{+}_{j}=\frac{i}{2}\phi_{j}(1-\cos\theta)\chi^{+}-\frac{1}{2}e_{+}(\theta_{j}+i\phi_{j}\sin\theta)\chi^{-},\label{A2}\\
&\chi^{-}_{j}=\frac{1}{2}e_{-}(\theta_{j}-i\phi_{j}\sin\theta)\chi^{+}-\frac{i}{2}\phi_{j} (1-\cos\theta)\chi^{-}\label{A3},
\end{align}
where the subscript $j$ denotes partial differentiation.

Keeping only the dominant contributions, as in Eqs.~(\ref{6}) and (\ref{7}), the Laplacian in (\ref{A1}) reads
\begin{align}\label{A4}
&\nabla^{2}\Psi=(\alpha^{(3)}_{xx}+\alpha^{(3)}_{yy}+\alpha^{(3)}_{zz})\chi^{+}+\\
&\left\{(\beta^{(3)}_{xx}+\beta^{(3)}_{yy}+\beta^{(3)}_{zz})-e_{+}\sum_{j=1}^{3}\alpha^{(3)}_{j}(\theta_{j}+i\phi_{j}\sin\theta)\right\}\chi^{-}\nonumber.
\end{align}
Thus, in WKB approximation the spatial wave functions satisfy
\begin{align}
\nabla^{2}\alpha^{(3)}+k^{2}_{+}\alpha^{(3)}&=0,\label{A5}\\
\nabla^{2}\beta^{(3)}+k^{2}_{-}\beta^{(3)}&=e_{+}\sum_{j=x,y,x}\alpha^{(3)}_{j}(\theta_{j}+i\phi_{j}\sin\theta),\label{A6}
\end{align}
where
\begin{equation}\label{A7}
k^{2}_{\pm}(x,y,z)=\frac{2m}{\hbar^{2}}\left[E-mgz\mp |\mu|B(x,y,z) \right]
\end{equation}
are the squared local wave numbers for the ($+$) and ($-$) spin state, respectively.

Now we consider a UCN with spin ($+$) starting at time $t=0$ at a TP and arriving at $t=\Delta T$ at the next TP which we label $U$. At $U$ the UCN momentarily moves along the local ES and we introduce a local Cartesian system of coordinates centered at $U$ with $x^{\prime}$ and $y^{\prime}$ in the plane of this ES. $z^{\prime}$ points away from the direction into which the UCN is reflected.\footnote{There are special cases where the path curvature at a TP equals the curvature of the ES. These are locations where two TPs coincide and the trajectory may proceed on either side of the ES, depending on the exact initial conditions. In this limit, the direction ``away'' is ill-defined, but for a continuous spectral distribution in phase space these paths represent a negligible fraction of the ensemble.} Coordinate system $x^{\prime}$, $y^{\prime}$, $z^{\prime}$ is defined for the narrow space where the particle motion can be considered linear and uniform.

Since the $\alpha^{(3)}$ and $\beta^{(3)}$ constituents of the wave function move as a unit the wave numbers $k_{x^{\prime}}$ and $k_{y^{\prime}}$ are the same for both. Thus we put
\begin{align}\label{A8}
&\alpha^{(3)}(x^{\prime},y^{\prime},z^{\prime})=\alpha(z^{\prime})\,e^{ik_{x^{\prime}}x^{\prime}}\,e^{ik_{y^{\prime}}y^{\prime}},\nonumber\\
&\beta^{(3)}(x^{\prime},y^{\prime},z^{\prime})=\beta(z^{\prime})\,e_{+}\,e^{ik_{x^{\prime}}x^{\prime}}\,e^{ik_{y^{\prime}}y^{\prime}}.
\end{align}
As in \cite{STE01}, $e_{+}=e^{i\phi}$ can be interpreted as a Bloch-wave modulation due to the field rotation.

Substituting (\ref{A8}) in (\ref{A5}) and (\ref{A6}) we obtain
\begin{equation}\label{A9}
\frac{d^{2}\alpha(z^{\prime})}{d z^{\prime 2}}+k^{\prime 2}_{+}(z^{\prime})\alpha(z^{\prime})=0
\end{equation}
and 
\begin{align}\label{A10}
&\left[\frac{d^{2}\beta(z^{\prime})}{d z^{\prime 2}}+k^{\prime 2}_{-}\beta(z^{\prime})\right]e^{ik_{x^{\prime}}x^{\prime}}\,e^{ik_{y^{\prime}}y^{\prime}}\nonumber\\
&=\sum_{j=x^{\prime}\textrm{,}y^{\prime}\textrm{,}x^{\prime}}\alpha^{(3)}_{j}(x^{\prime},y^{\prime},z^{\prime})(\theta_{j}+i\phi_{j}\sin\theta),
\end{align}
where the wave numbers for the $z^{\prime}$ direction are given by
\begin{equation}\label{A11}
k^{\prime 2}_{\pm}=k^{2}_{\pm}-k^{2}_{x^{\prime}}-k^{2}_{y^{\prime}}
\end{equation}
with $k^{2}_{\pm}$ defined in (\ref{A7}). Among the components of $k_{+}$ and $k_{-}$, the $z^{\prime}$ component $k^{\prime}_{+}(z^{\prime})$ plays a special role. Even within the narrow space where the primed system of coordinates has been defined, $k^{\prime}_{+}$ is not constant. It becomes zero at the TP $z^{\prime}=0$ and, in this semi-classical picture, is only defined for $z^{\prime}\le 0$.

In WKB approximation the solution of (\ref{A9}) is
\begin{equation}\label{A12}
\alpha(z^{\prime})=\frac{1}{\sqrt{k^{\prime}_{+}(z^{\prime})}}\,e^{iX^{\prime}_{+}(z^{\prime})}
\end{equation}
with $X^{\prime}_{+}(z^{\prime})=\int^{z^{\prime}}k^{\prime}_{+}(u)du$. (The exact quantum  solution in form of an Airy function has no singularity at $z^{\prime}=0$ and decays exponentially in the classically forbidden zone $z^{\prime}>0$.)

To solve (\ref{A10}) we substitute the WKB approximation for the partial derivatives on the RHS, $\alpha^{(3)}_{x^{\prime}}=ik_{x^{\prime}}\alpha^{(3)}$, $\alpha^{(3)}_{y^{\prime}}=ik_{y^{\prime}}\alpha^{(3)}$, $\alpha^{(3)}_{z^{\prime}}=ik^{\prime}_{+}(z^{\prime})\alpha^{(3)}$, and implement the total time derivative in the form $d/dt=v_{x^{\prime}}(\partial/\partial x^{\prime})+v_{y^{\prime}}(\partial/\partial y^{\prime})+v_{z^{\prime}}(\partial/\partial z^{\prime})$ with velocity $\mathbf{v}=(\hbar/m)\mathbf{k}_{+}$. Employing also (\ref{A12}), (\ref{A10}) becomes
\begin{equation}\label{A13}
\frac{d^{2}\beta(z^{\prime})}{dz^{\prime 2}}+k^{\prime 2}_{-}\beta(z^{\prime})=\frac{m}{\hbar}\frac{i}{\sqrt{k^{\prime}_{+}(z^{\prime})}}(\dot{\theta}+i\dot{\phi}\sin\theta)\,e^{iX^{\prime}_{+}(z^{\prime})}.
\end{equation}

The solution of (\ref{A13}) has been outlined in \cite{WAL01,STE01}. The phase factor for the wave $\beta(z^{\prime})$ is the same as for $\alpha(z^{\prime})$: $e^{i X^{\prime}_{+}}(z^{\prime})$. Therefore, in WKB approximation we have $d^{2}\beta(z^{\prime})/dz^{\prime 2}=-k^{\prime 2}_{+}(z^{\prime})\beta(z^{\prime})$ and the solution of (\ref{A13}) becomes
\begin{equation}\label{A14}
\beta(z^{\prime})=\frac{m}{\hbar}\frac{i}{\sqrt{k^{\prime}_{+}(z^{\prime})}}\frac{\dot{\theta}+i\dot{\phi}\sin\theta}{k^{\prime 2}_{-}-k^{\prime 2}_{+}(z^{\prime})}e^{iX^{\prime}_{+}(z^{\prime})}.
\end{equation}

The depolarization loss measured at TP $U$ is given by the probability current for spin-flipped UCNs,
\begin{equation}\label{A15}
j_{-}(z^{\prime})=-\frac{\hbar}{m}\real\left[i\beta^{\ast}(z^{\prime})\left(\frac{d\beta}{dz^{\prime}} \right) \right],
\end{equation}
leaving the storage space at $z^{\prime}=0$ in the positive $z^{\prime}$ direction. With (\ref{A14}) this current is
\begin{equation}\label{A16}
j_{-}(z^{\prime})=\frac{m}{\hbar}\frac{\dot{\theta}^{2}+\dot{\phi}^{2}\sin^{2}\theta}{[k^{\prime 2}_{-}-k^{\prime 2}_{+}(z^{\prime})]^{2}}=\frac{\hbar}{m}\frac{\Omega^{2}}{4\omega^{2}_{L}},
\end{equation}
evaluated at $z^{\prime}=0$ (i.e.,~for the field variables $\Omega$ and $\omega_{L}$ at the particle position at time $t=\Delta T$). In the last step of (\ref{A16}) we have used the Larmor frequency $\omega_{L}=\hbar (k^{\prime 2}_{-}-k^{\prime 2}_{+})/(2 m)$.

To evaluate the spin-flip loss rate $1/\tau_{dep}$ between the consecutive TPs we divide the current (\ref{A16}) by the number $\mathcal{N}$ of ($+$) spin UCNs moving between the TPs in a channel with unit cross section centered at the trajectory. The cross section of this channel is measured parallel to the ES at every point along the path. The $\mathcal{N}$ particles within this volume contribute to loss current (\ref{A16}), their decay rate $-\dot{\mathcal{N}}$ equaling $j_{-}(0)$.

Denoting the wave number perpendicular to the ESs traversed along the way by $k^{\prime}_{+}(t)$ and using the WKB form $|\alpha(t)|^{2}=1/k^{\prime}_{+}(t)$ as the particle density we have
\begin{align}\label{A17}
\mathcal{N}&=\int_{\textrm{channel}}|\alpha(t)|^{2}\,d(volume)\nonumber\\
&=\int_{0}^{\Delta T}\frac{1}{k^{\prime}_{+}(t)}\frac{\hbar k^{\prime}_{+}(t)\,dt}{m}=\frac{\hbar}{m}\Delta T.
\end{align}
As for the 1D field model of \cite{STE01}, $\mathcal{N}$ is given directly by the travel time $\Delta T$. Using (\ref{A16}), the depolarization rate becomes
\begin{equation}\label{A18}
1/\tau_{dep}=-\dot{\mathcal{N}}/\mathcal{N}=\frac{m}{\hbar}\frac{j_{-}(0)}{\Delta T}=\frac{\Omega^{2}}{4\omega^{2}_{L}\,\Delta T},
\end{equation}
evaluated for the field at the endpoint $U$. This agrees with the semi-classical result $1/\tau_{dep}=p(\Delta T)/\Delta T$ with $p(t)$ given by Eq.~(\ref{10}). Generalizing this result to arbitrary time $t$, we choose the UCN position at $t$ as the center of reference system $x^{\prime},y^{\prime},z^{\prime}$, with $z^{\prime}$ normal to the local ES, and use (\ref{A14}), (\ref{A15}) to obtain the identity
\begin{equation}\label{A19}
\frac{m}{\hbar} j_{-}(t)=\frac{\Omega^{2}(t)}{4\omega^{2}_{L}(t)}=p(t).
\end{equation}
This shows that the semi-classical and the quantum approaches to depolarization are equivalent, with $p(t)$ directly corresponding to $m/\hbar$ times the probability current $j_{-}(t)$.   

There is an open question of interpretation: In the derivation of (\ref{A13}) we used a total time derivative in the form $\dot{f}=v_{x^{\prime}}(\partial f/\partial x^{\prime})+v_{y^{\prime}}(\partial f/\partial y^{\prime})+v_{x^{\prime}}(\partial f/\partial z^{\prime})$ with velocity $\mathbf{v}$ referring to the particle's motion along its classical path. In this sense, the quantum approach of this Appendix does involve classical concepts. Use of the WKB method is not the only approximation made.

A similar caveat applies to the possibility of going to higher-order approximations in this quantum approach. In the semi-classical analysis we were allowed to add, in (\ref{21}) and (\ref{22}), terms such as $A_{pp}\beta$ which are of second order and had been neglected in the first-order approach of Eqs.~(\ref{6}) and (\ref{7}). However, adding the corresponding second-order contributions in the quantum treatment would require that we also add the second-order quantities neglected in the WKB approximation used to derive the ODEs (\ref{A5}) and (\ref{A6}). These would be replaced by coupled non-linear PDEs of high complexity. At this stage the semi-classical and quantum approaches clearly diverge.


\begin{thebibliography}{100}
 
\bibitem{BHA01} T.~Bhattacharya, V.~Cirigliano, S.~D.~Cohen \textit{et al.}, Phys.~Rev.~D $\mathbf{85}$, 054512 (2012).
\bibitem{CIR01} V.~Cirigliano, S.~Gardner and B.~R.~Holstein, Progress in Particle and Nuclear Physics $\mathbf{71}$, 93 (2013).
\bibitem{GAR01} S.~Gardner and B.~Plaster, Phys.~Rev.~C $\mathbf{87}$, 065504 (2013).
\bibitem{MEN01} G.~Mention, M.~Fechner, Th.~Lasserre \textit{et al.}, Phys.~Rev.~D $\mathbf{83}$, 073006 (2011).
\bibitem{ZHA01} C.~Zhang, X.~Qian and P.~Vogel, Phys.~Rev.~D $\mathbf{87}$, 073018 (2013).
\bibitem{COC01} A.~Coc, J.-P.~Uzan and E.~Vangioni, J.~Cosmology and Astroparticle Physics $\mathbf{10}$, 050 (2014).
\bibitem{IOC01} F.~Iocco, G.~Mangano, G.~Miele \textit{et al.}, Physics Reports $\mathbf{472}$, 1 (2009).
\bibitem{MAT01} G.~J.~Mathews, T.~Kajino and T.~Shima, Phys.~Rev.~D $\mathbf{71}$, 021302 (2005).
\bibitem{ABE01} H.~Abele, Prog.~Part.~Nucl.~Phys.~$\mathbf{60}$, 1 (2008).
\bibitem{NIC01} J.~S.~Nico, J.~Phys.~G: Nuclear and Particle Physics $\mathbf{36}$, 104001 (2008).
\bibitem{PAU01} S.~Paul, Nucl.~Instr.~Meth.~Phys.~Res.~A $\mathbf{611}$, 157 (2009).
\bibitem{DUB01} D.~Dubbers and M.~G.~Schmidt, Rev.~Mod.~Phys.~$\mathbf{83}$, 1111 (2011).
\bibitem{WIE01} F.~E.~Wietfeldt and G.~L.~Greene, Rev.~Mod.~Phys.~$\mathbf{83}$, 1173 (2011).
\bibitem{SEE01} S.~J.~Seestrom, ed., \textit{Next generation experiments to measure the neutron lifetime}, Proc.~Workshop Santa Fe, Nov.~2012, World Scientific, 2014.
\bibitem{VLA01} V.~V.~Vladimirsky, JETP $\textbf{12}$, 740 (1961).
\bibitem{POK01} Yu.~N.~Pokotilovski, JETP Lett.~$\textbf{76}$, 131 (2002); $\textit{Erratum}$, JETP Lett.~$\textbf{78}$, 422 (2003).
\bibitem{MAJ01} E.~Majorana, Il Nuovo Cimento $\mathbf{9}$, 43 (1932).
\bibitem{WAL01} P.~L.~Walstrom, J.~D.~Bowman, S.~I.~Penttila \textit{et al.}, Nucl. Instr. Methods Phys. Res. A $\mathbf{599}$, 82 (2009).
\bibitem{STE01} A.~Steyerl, C.~Kaufman, G.~M\"uller, S.~S.~Malik, A.~M.~Desai, Phys. Rev. C $\mathbf{86}$, 065501 (2012).
\bibitem{STE02} A.~Steyerl, C.~Kaufman, G.~M\"uller, S.~S.~Malik, A.~M.~Desai, in \textit{2012 Next generation experiments to measure the neutron lifetime}, ed.~ S.~J.~Seestrom, World Scientific (2014), pp.~75-86.
\bibitem{HUF01} P.~Huffman, C.~R.~Brome, J.~S.~Butterworth \textit{et al.}, Nature $\mathbf{403}$, 62 (2000).
\bibitem{EZH01} V.~F.~Ezhov, A.~Z.~Andreev, G.~Ban \textit{et al.}, Nucl.~Instr.~Methods Phys.~Res.~A $\mathbf{611}$, 167 (2009); \textit{Measurement of the neutron lifetime with ultracold neutrons stored in a magneto-gravitational trap}, arXiv.org/pdf/1412.7434.
\bibitem{LEU01} K.~Leung and O.~Zimmer, Nucl.~Instr.~Methods Phys.~Res.~A $\mathbf{611}$, 181 (2009).
\bibitem{LEU02} K.~K.~H.~Leung, P.~Geltenbort, S.~Ivanov, F.~Rosenau, O.~Zimmer, Phys.~Rev.~C $\mathbf{94}$, 045502 (2016).
\bibitem{BEC01} M.~Beck, K.~Eberhardt, Ch.~Geppert \textit{et al.}, in \textit{International workshop: Probing fundamental symmetries and interactions with UCN}, April 2016, https://indico.mitp.uni-mainz.de/event/59.
\bibitem{MAT02} S.~Materne, R.~Picker, I.~Altarev \textit{et al.}, Nucl.~Instr.~Methods Phys.~Res.~A $\mathbf{611}$, 176 (2009).
\bibitem{PAU02} W.~Paul, F.~Anton, L.~Paul, S.~Paul, W.~Mampe, Z.~Physik C $\mathbf{45}$, 25 (1989).
\bibitem{SAL01} D.~J.~Salvat, E.~R.~Adamek, D.~Barlow \textit{et al.}, Phys.~Rev.~C $\mathbf{89}$, 052501 (2014).
\bibitem{SAU01} A.~Saunders, M.~Makela, Y.~Bagdasarova \textit{et al.}, Rev.~Sci.~Instrum.~84, 013304 (2013).
\bibitem{MOR01} P.~M.~Morse and H.~Feshbach, \textit{Methods of theoretical physics}, McGraw Hill, New York, 1953, Chap.~9.3.
\bibitem{GOL01} H. Goldstein, \textit{Classical mechanics}, Addison-Wesley, London, 1950.
\bibitem{BRO01} C.~R.~Brome, J.~S.~Butterworth, S.~N.~Dzhosyuk \textit{et al.}, Phys.~Rev.~C $\mathbf{63}$, 055502 (2001).
\bibitem{YAN01} L.~Yang, C.~R.~Brome, J.~S.~Butterworth \textit{et al.}, Rev.~Sci.~Instr.~$\mathbf{79}$, 031301 (2008).
\bibitem{STE03} A.~Steyerl, C.~Kaufman, G.~M\"uller \textit{et al.}, Phys. Rev. A $\mathbf{89}$, 052129 (2014).
\bibitem{GRE01} G.~L.~Greene and P.~Geltenbort, \textit{The neutron enigma}, Sci.~American, April 2016.

\end{thebibliography}
\end{document}